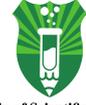

Deanship of Scientific Research

# The Repercussions of the COVID-19 Pandemic on Higher Education and its implications for Syrian Refugees Students (An Analytical Descriptive Study)

*Anas Alsobeh\*, Ahlam Aloudat*

Yarmouk University, Jordan.

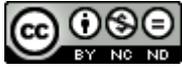

**Abstract**

This study aims to reveal the most important challenges and difficulties that refugee students faced in Jordanian universities (e.g., Yarmouk University, AL Al-Bayt, and the Private Zarqa University) due to the COVID-19 pandemic through measuring a different of indicators that are related, in addition, to identify some of the independent variables on e-educational challenges. In the study, the analytical description approach was used. The data collection tool is a questionnaire, which was distributed to a random sample of students electronically. Results show that the necessity to implement educational and psychological counseling programs and economic support programs to support the e-Learning costs. The study confirmed that refugees are the most affected students with the pandemic compared to the host community.

**أثر جائحة كورونا في التعليم العالي و انعكاساته لدى طلبة الجامعات الأردنية من اللاجئين السوريين (دراسة وصفية تحليلية)**

*أنس الصبح\*، أحلام العودات*

جامعة اليرموك، الأردن.

**ملخّص**

هدفت الدراسة الحالية إلى الكشف عن أهم التحديات والصعوبات التي تواجه طلبة الجامعات الأردنية من اللاجئين السوريين (جامعة اليرموك، آل البيت، جامعة الزرقاء الخاصة) نتيجة أعباء الجائحة عن طريق قياس عدد من المؤشرات ذات العلاقة بالدراسة، فضلا عن كشف أثر بعض المتغيرات المستقلة في تلك التحديات والصعوبات وذلك بالاعتماد على المنهج الوصف التحليلي وباستخدام عدة أساليب إحصائية. وتمثلت أداة جمع البيانات في استبانة أعدت خصيصا لتحقيق أهداف الدراسة جرى توزيعها على عينة عشوائية من الطلبة إلكترونيا. وتوصلت الدراسة لعدد من النتائج وأهمها ضرورة تنفيذ برامج تعليمية إرشادية ونفسية وبرامج دعم اقتصادية لتوفير تكاليف الدراسة الناجمة من التعلم عن بعد، وأكدت الدراسة أن اللاجئين أكثر تأثرا في الجائحة مقارنة بأفراد المجتمع المضيف.

**الكلمات الدالة**: اللاجئين السوريين، جائحة كورونا، التعلم عن بعد.

## المقدمة:

بدأت الأزمة السورية منذ أكثر من عشر سنوات، وبالتحديد في عام (2011)، واستمرت إلى يومنا هذا، حيث لجأ مئات الآلاف من السوريين إلى الأردن على شكل هجرات قسرية، مُلقية بظلالها بما حملته من تداعيات أثرت في الأردن من النواحي الاقتصادية والاجتماعية، وفي ذات الوقت خلفت العديد من الآثار السلبية على الملايين من السوريين.

ويستضيف الأردن حاليًا أكثر من مليون لاجئ سوري، منهم حوالي (665.000) لاجئ فقط مسجلين لدى المفوضية السامية للأمم المتحدة لشؤون اللاجئين. ويقيم ما يقرب (10٪) من هؤلاء اللاجئين في المخيمات المخصصة لهم، في حين يسكن البقية داخل المجتمعات الحضرية والريفية في الأردن، وقد تكون الأزمة السورية من أكثر القضايا تعقيدًا وطول الأمد التي واجهتها المملكة في العقود الأخيرة، سواء بسبب تكاليفها الإنسانية والاجتماعية والاقتصادية والبيئية التي لا مفر منها، أو تأثيرها المباشر على قدرة الأردن على الحفاظ على مكاسبه التنموية، والحفاظ على نتائج الإصلاحات وبرامج التكيف الاقتصادية، والاستجابة للتحديات المقبلة (سميران وسميران، 2014).

وانطلاقًا من التزام الأردن تجاه قضايا اللجوء والهجرة، فقد سعت الحكومة الأردنية دائمًا إلى منح اللاجئين السوريين معاملة متساوية أو حتى تفضيلية في الحصول على الخدمات التي تضمن الحد الملائم من المستوى المعيشي إلى حين العودة إلى بلادهم، ولا سيما في مجالات التعليم والصحة والمرافق العامة. ومع ازدياد طول الأزمة، أصبحت القدرة الاستيعابية للأردن مشبعة؛ استنفدت خلالها جميع آليات في التكيف مع الأزمة ووصلت الحد الأقصى (الجراح، 2018).

يزيد عدد الطلاب السوريين في الأردن عن (126) ألف طالب وطالبة، حسب أرقام وزارة التربية الأردنية، في بداية العام الدراسي (2018-2019)، بمن فيهم طلاب مخيمات اللجوء، ويقوم بتدريسهم حوالي (5800) معلمًا ومعلمة، يحملون شهادات جامعية في مختلف التخصصات التعليمية، يتوزعون على (48) مدرسة في مخيمي الزعتري والأزرق، ونحو (200) مدرسة منتشرة في جميع المدن الأردنية، تعمل بنظام الفترتين، يتركز معظمها في المدن الشمالية، الرمثا، إربد، المفرق، حيث كثافة الوجود السوري هي الأكبر، ويتمتع الطلاب السوريون في المدارس الحكومية الأردنية، بالإعفاء من أية رسوم أو تكاليف مادية، سواء كانت رسومًا مدرسية أو ثمن كتب (وزارة التعليم العالي والبحث العلمي، 2019).

أما على صعيد التعليم الجامعي، فيبلغ عدد الطلبة من اللاجئين السوريين المسجلين في الجامعات الأردنية مع بداية العام الدراسي (2018- 2019) نحو (15) ألف طالب وطالبة، حسب دائرة الإحصاءات في وزارة التعليم العالي الأردنية (وزارة التعليم العالي، 2019).

وقد لقي تعليم اللاجئين السوريين في الأردن، اهتمامًا منذ بداية أزمة اللجوء السوري، من قبل الحكومة الأردنية، والمنظمات الدولية، وحتى من اللاجئين أنفسهم، ولكن مع طول سنوات اللجوء برزت بعض الصعوبات في العملية التعليمية.

وأوجدت جائحة كورونا (كوفيد – 19) أكبر انقطاع في نظم التعليم بمختلف المراحل الأساسية والثانوية والجامعية في التاريخ، وهو ما تضرر منه نحو (1.6) بليون من الطلبة وفي جميع القارات. وأثرت في أكثر من (190) دولة من خلال عمليات إغلاق المدارس وغيرها من أماكن التعليم في العالم، وهي نسبة ترتفع لتصل إلى (99٪) في البلدان المنخفضة الدخل والبلدان المتوسطة الدخل الأمر الذي فاقم أزمة الفوارق التعليمية القائمة على الحد من فرص الكثير من الأطفال والشباب والبالغين المنتمين إلى أشد الفئات ضعفًا – أولئك الذين يعيشون في مناطق فقيرة أو ريفية والفتيات واللاجئون، والأشخاص ذوي الإعاقة والمشردين قسرًا)، وثمة تخوف من تسرب نحو (22.5) مليون طفل من النظام التعليمي من مرحلة التعليم الابتدائي من المدارس في العام المقبل بسبب التأثير الاقتصادي للجائحة (Save the Children, 2020).

وتأثر النظام التعليمي الأساسي والتعليم الجامعي في الأردن شأنه شأن بقية دول العالم، من جوانب عديدة، فقد عمدت المدارس والجامعات إلى الإغلاق واتباع نظام التعلم الإلكتروني عن بُعد، مما ألقى بظلاله وانعكاسات سيئة على المجالات النفسية، والمجالات الأكاديمية، والمجالات الاقتصادية، وخاصة على اللاجئين السوريين لما لهم من خصوصية العيش في الأردن.

## مشكلة الدراسة:

تكمن مشكلة الدراسة في تعرُّف الآثار الناتجة عن أزمة كورونا على الطلبة اللاجئين من السوريين في الجامعات الأردنية موزعة على الأقاليم الثلاث الشمال والوسط والجنوب. تستند الدراسة إلى تعرُّف أثر جائحة كورونا المستجد في الطلبة، وكيفية تأقلمهم مع التغيرات التي طرأت على التعليم بالجامعات الأردنية وتحديد الدروس المستنبطة والنتائج الإيجابية المحتملة للحظر العالمي على التعاليم العالي وأثرها المستمر والدائم في التدريس في الجامعات الأردنية وكيفية تعزيز النظام التعليمي وتغيراته.

## أسئلة الدراسة:

وتسعى الدراسة إلى الإجابة عن الأسئلة الآتية:

1. ما أثر انعكاسات جائحة كورونا في الطلبة السوريين الدارسين في الجامعات الأردنية؟
2. هل هناك فروق ذات دلالة إحصائية عند مستوى الدلالة ($\alpha \leq 0.05$) بين تقديرات أفراد العينة لانعكاسات جائحة كورونا على الطلبة

السوريين الدارسين في الجامعات الأردنية تعزى لمتغيرات الجنس، والحالة الاجتماعية، والفئة العمرية، والإقليم، ومكان السكن، والمستوى الدراسي، والمعدل التراكمي، وصفة الجامعة، وهل تعمل؟

**أهمية الدراسة:**

**الأهمية العلمية:** تأتي أهمية هذه الدراسة في أن جائحة كورونا تعدُّ من الظروف الطارئة الاستثنائية التي يمر بها العالم أجمع، وأن أي دراسة في هذا المجال وفي ظل هذه الظروف تعد إضافة نوعية جديدة للدراسات والبحوث التي تعني بتعرُّف انعكاسات الأزمات العالمية على البشرية وتحديدًا المستضعفين منهم في مجالات الحياة المختلفة، ولذا تأتي هذه الدراسة انموذجا لدراسة التحديات الناجمة عن الجائحة في التعليم العالي و بما يخص اللاجئين بالاردن.

**الأهمية العملية:** تأتي أهمية هذه الدراسة من خلال النتائج التي أسفرت عنها لتكون دليلًا ومرجعًا لبرامج الإغاثة والمساعدات التي سيتم استحداثها للتخفيف عن الطلبة التعليم العالي من اللاجئين السوريين في الجامعات الأردنية، سواء من المنظمات غير الحكومية (NGOs)، أو الحكومات العربية والأجنبية.

**أهداف الدراسة**

هدفت هذه الدراسة إلى تعرُّف انعكاس جائحة كورونا على الطلبة اللاجئين السوريين الدارسين في الجامعات الأردنية وذلك من خلال:

1. تعرُّف انعكاسات جائحة كورونا على الطلبة السوريين.
2. توضيح انعكاس جائحة كورونا على المتغيرات الديموغرافية لعينة الدراسة.
3. تعرُّف انعكاس جائحة كورونا على ابعاد الدراسة والمتمثلة في البعد الأكاديمي، الاجتماعي، النفسي والاقتصادي الدراسة.

**حدود الدراسة**

**الحدود الزمنية:** تقتصر هذه الدراسة على الفترة الزمنية من بداية جائحة كرونا (شهر آذار 2020 وحتى شهر أيلول 2020).

**الحدود المكانية:** تم تطبيق هذه الدراسة على الطلبة اللاجئين السوريين ممن هم على مقاعد الدراسة في الجامعات الأردنية.

**حدود بشرية:** الطلبة السوريين المسجلين في الجامعات الأردنية، خلال العام الجامعي (2015 – 2016) وما بعده.

**التعريفات الإجرائية**

**انعكاسات:**

يُعرف الانعكاس على أنه اِرتِدَادٌ، أو أَثَرٌ، أو اِنْقِلَابٌ (أبادي، 1985). وتُعرف اجرائيًا على أنها مجموعة المجالات فقراتها المُعدة في الاستبانة لأغراض هذه الدراسة.

**جائحة كورونا:** الجائحة (Pandemic) تحدث عندما ينتشر الوباء إلى عدة بلدان أو قارات، وعادة ما يصاب عدد كبير من سكان الكرة الأرضية بهذا الوباء، في ظل عدم وجود لقاحات لمقاومته. وهي جائحةٌ عالميةٌ مستمرةٌ حاليًا لمرض فيروس كورونا (كوفيد-19)، سببها فيروس كورونا المرتبط بالمتلازمة التنفسية الحادة الشديدة (سارس-كوف-2)، تفشّى المرض للمرة الأولى في مدينة ووهان الصينية في أوائل شهر ديسمبر عام 2019، حيث أعلنت منظمة الصحة العالمية رسميًا في 30 يناير أن تفشي الفيروس يُشكل حالة طوارئ صحية عامة تبعث على القلق الدولي (UNFPA, 2020).

**الطلبة السوريين الدارسين في الجامعات الأردنية:** هم الطلبة الذين تم قبولهم في الجامعات الأردنية نتيجة الأحداث في الجمهورية العربية السورية خلال العام الجامعي (2015 – 2016) وما بعده.

**الدراسات السابقة**

اطّلع الباحثان على الدراسات السابقة التي لها علاقة بموضوع الدراسة الحالية، ووجد عدة دراسات، وتم عرضها من الأحدث إلى الأقدم على النحو الآتي:

أجرت ملكاوي (2020) دراسة بعنوان: "التعلم عن بعد واقع وتحديات من وجهة نظر أولياء الأمور خلال جائحة فيروس كورونا "كوفيد19" في محافظة اربد في الأردن" هدفت إلى تعرُّف واقع التعلم عن بعد وتحدياته خلال جائحة فيروس كورونا"كوفيد19"من وجهة نظر أولياء الأمور في محافظة إربد في الأردن. استخدمت الدراسة المنهج الوصفي التحليلي، وتم جمع المعلومات من خلال استبانة، بلغت عينتها (953) من أولياء الأمور. تكونت الاستبانة من 20 فقرة موزعة على مجالين: واقع التعلم عن بعد وتحدياته خلال جائحة فيروس كورونا "كوفيد19"من وجهة نظر أولياء الأمور في محافظة اربد، في الأردن، والمجال الثاني آليات مواجهة تحديات التعلم عن بعد خلال جائحة فيروس كورونا "كوفيد19"من وجهة نظر أولياء الأمور في محافظة إربد، في الأردن. توصلت الدراسة إلى النتائج: أن واقع التعلم عن بعد وتحدياته خلال جائحة فيروس كورونا"كوفيد19 "من وجهة نظر أولياء الأمور جاءت بدرجة متوسطة، وجاء مجال الدراسة - واقع التعلم عن بعد وتحدياته خلال جائحة فيروس كورونا"كوفيد19"من وجهة نظر أولياء الأمور بدرجة متوسطة، أما المجال الثاني، آليات مواجهة تحديات التعلم عن بعد خلال جائحة فيروس، كما لا توجد فروق ذات دلالة إحصائية تُعزى لمتغير صفة ولي الأمر.

152

وأجرى ناصيف وكوموش (2020) دراسة بعنوان: "حتمية العمل عن بعد والتحول نحو الإدارة الالكترونية في منظمات المجتمع المدني السورية في ظل كورونا: دراسة تطبيقية على إدارة المنظمات السورية العاملة في تركيا من تركيا) " حيث يشهد العالم جائحة كورونا في عصرنا الحالي تحول العمل في الأقسام الإدارية للمنظمات السورية العاملة من تركيا الى نظام العمل عن بعد وظهر معه الحاجة إلى الانتقال لنظام الإدارة الإلكترونية. كشفت الدراسة ان المنظمات السورية تملك المقومات الأزمة لتحول نحو الإدارة الإلكترونية وتحتاج لفترة زمنية تمتد حتى سنتين للتحول، وتواجه بعض الضعف في بعض المتطلبات، حيث اوصت الدراسة المسؤولين في المنظمات بضرورة العمل على تلافيها، كما أكدت على ضرورة الاهتمام بتطبيق الإدارة الإلكترونية لما لها من فوائد على العاملين في أثناء تأديتهم أعمالهم، وكذلك في فترات العمل عن بعد للحفاظ على أداء عام جيد للمنظمة.

كما أجرى رومي وهانسنسن وهانسنسن (Romi, Hansenson & Hensenson, 2019) دراسة هدفت إلى تعرُّف اتجاهات الطلبة في الصف العاشر نحو التعليم الإلكتروني وأثر المستوى الاقتصادي والاجتماعي والقدرة في استخدام الحاسب الآلي. تكونت عينة الدراسة من مجموعتين من(60) طالبًا موزعين بالتساوي على مجموعتين، الأولى تلقت تعليمًا وجاهيًا، والثانية تلقت تعليمًا باستخدام الحاسب الآلي والتعليم الإلكتروني. توصلت النتائج إلى عدم وجود فروق للمتغيرات الديمغرافية على الاتجاهات، وأن اتجاهات طلبة المجموعة الأولى تجاه التعليم الإلكتروني كانت أعلى من اتجاهات طلبة المجموعة الثانية, ووجود علاقة بين القدرة على استخدام الحاسب الآلي والتعليم الإلكتروني وبين الاتجاهات الموجبة نحوه.

في دراسة الجراح (2018) بعنوان: "تحليل مخاطر اللجوء على الأمن الإنساني: دراسة حالة أزمة اللجوء السوري في الأردن. تكونت عينة الدراسة من (125) مبحوثًا جرى اختيارهم بطريقة عشوائية. استخدم المنهج الوصفي التحليلي بطريقتي المسح ودراسة الحالة. وتم استخدام استبانة كأداة لجمع البيانات. توصلت الدراسة إلى أن (18.6%) من المخاطر التي تم تحديدها في التي يمكن أن تنتج عن أزمة اللجوء السوري وتبعاتها في الأردن، خلال السنوات الخمس القادمة، جاءت بمستوى خطورة "مرتفع جدًا". وأن (79%) منها جاءت بمستوى خطورة "مرتفع".

كما أجرى اركورفل وابيدو واركوفل ونيلي (Arkorful, Abaidoo, Arkorful & Nelly,2018) دراسة هدفت إلى الكشف عن فاعلية استخدام التعلم الالكتروني في التدريس في مؤسسات التعليم العالي. واستعرضت الدراسة الأدب النظري وغطت خلفية علمية للدراسة من خلال مراجعة المساهمات التي قدمها العديد من الباحثين والمؤسسات حول مفهوم التعلم الالكتروني، وخاصة استخدامه في التدريس والتعلم في مؤسسات التعليم العالي. تكشف النقاب عن بعض وجهات النظر التي تبادلها الناس والمؤسسات عالميًا حول اعتماد ودمج تقنيات التعلم الالكتروني في التعليم مين خلال الاستقصاءات والملاحظات الاخرى. تبحي في معنى أو تعريف التعلم الالكتروني كما قدمه باحثون مختلفون والدور الذي يلعبه التعلم الالكتروني في مؤسسات التعليم العالي في ما يتعلق بعمليات التدريس والتعلم، ومزايا وعيوب اعتماده وتنفيذه.

كما قدم الباحث (عدنان الجادري، 2020) بدراسة بعنوان "الانطلاق نحو جامعات بحثية :جامعة عمان العربية نموذجًا " اثر تحول الجامعات من النمط التدريسي إلى النمط البحثي وتأتي اهمية هذة الدراسة بحث دورالبحث العلمي في الحياة الاقتصادية والاجتماعية والثقافية وتبوؤها المكانة الريادية الفاعلة في هذا المضمار. وتاتي هذة الاهمية من امتلاك الجامعات الاردنية الإمكانات العلمية والبشرية والفنية بدرجة عالية الاداء .وبالنهاية اكدت هذة الدراسة على ان الجامعات الاردنية بمفهومها المعاصر هي مصنع المعرفة وليست ترفاً ثقافياً وأصبحت خيارًا استراتيجيا في إطار منظومة استثمار البحثي والصناعي وتنمية الموارد البشرية لمواكبة التطور والتقدم.

قدم البنك الدولي دراسة أظهرت هذة الدراسة التحديات التي واجهت التعليم المدرسي والجامعي جراء تفشي فيروس كورونا. وبينت هذه الدراسة العجز والنقص بالبنية التحتية لاتاحة التعليم الإلكتروني نتيجة المعانة من نقص خدمات الكهرباء أو الإنترنت أو التفاز، بالإضافة إلى عدم قدرة الاباء والامهات اخذ دور المدرس لمساعدة ابنائهم. وعرضت هذة الدراسة مثال للتحديات التي زادت من عبىء الجائحة خلال الجائحة وهي "روزا معلمة في بيكاسي بإندونيسيا. وتدرس طفلها في مدرسة خاصة وتستطيع التعلّم عن طريق الإنترنت لساعات عديدة في اليوم. لكنها تجد أن الجمع بين وظيفتها ومسؤولياتها الأسرية، ورداءة الاتصال بالإنترنت يجعلان التدريس والتعلّم أكثر صعوبة في أثناء الجائحة". (البنك الدولي، 2021)

قدمت الباحثة لجوزيليا نيفيس بالمقالة " كوفيد-19 يقدم فرصة لإعادة النظر في أساليب الامتحانات التقليديه " فرص المكتسبة من التحول من التعلم التقليدي إلى التعليم الاكتروني حيث يسهم هذا التحول إلى تطوير مهارات الطلبة خلال القرن الحادي والعشرين في: حل المشكلات، والتفكير النقدي، والتواصل، والتعاون، والإبداع، والذكاء العاطفي، والعزيمة، والدافع الجوهري، والتفاهم الاجتماعي والثقافي. ومن ناحية اخرى تحول نمط الاختبارات التقليدية إلى الاكترونية ليتناسب مع التغيرات الجديدة والكفاءات المعرفية والشخصية الذاتية والشخصية مع الآخرين التي نريد أن يكتسبها الطلبة؟ كما قدم الباحث فرصًا لإعادة النظر في الممارسات التعليمية (والتقييمية) الحالية. يمكن لأنظمة التعلم الرقمية الحالية ببساطة دعم النشاطات التقليدية، وإدامة عقليات القرن التاسع عشر، أو تقريب التعليم من ذلك الواقع غير المادي المتصل على الانترنت دائمًا الذي نعيش فيه.

وأجرى الصندوق الائتماني الأوروبي (2018) دراسة تحت عنوان: "التحدي التعليمي واستثمار الفرص وتحديد العوائق في التعليم العالي للطلبة السوريين في الأردن". حيث هدف المشروع إلى توفير معرفة واسعة عن الطلبة السوريين في الجامعات الأردنية، وتحليل للوضع الحالي للطالب السوري المسجلين في مؤسسات التعليم العالي في الأردن. بالإضافة إلى البحث في القضايا الرئيسة التي يواجها الطلبة في متابعة تعليمهم والتحديات الرئيسة التي

153

واجهتهم أو قد تواجههم. تكون مجتمع الدراسة من الطلبة السوريين في الجامعات الأردنية الحكومية والخاصة، لتحقيق أهداف الدراسة تم اخذ عينة حجمها (1675) طالب وطالبة من (18) جامعة حكومية وخاصة. توصلت نتائج الدراسة إلى أن الغاية الأساسية من التحاق الطلبة السوريين في التعليم العالي تتمثل في الحصول على وظيفة أفضل في المستقبل، ومن ثم تكوين معرفة ذاتية وثقافة عالية، وقد ظهر هذا على نحو واضح في مجموعات النقاش المركزة، الذي أكد فيه الطلبة على اصرارهم في الصمود في وجه الظروف التي مروا فيها من خلال استكمال التعليم وتطوير ذاتهم ومعارفهم وخبراتهم في الحياة. كما بينت النتائج أن غالبية الطلبة اختاروا التخصص الذي يدرسونه بناءً على ميولهم ورغبتهم الشخصية، ويعتقدون بأن التخصص الذي اختاروه يحقق طموحاتهم المستقبلية. وأشارت النتائج أن الصعوبات التي واجهها الطلبة عند تقديم طلبات الالتحاق في الجامعة تخلص في معادلة شهاداتهم المدرسية، واحضار الأوراق الثبوتية المطلوبة من سورية، بالإضافة الى الإجراءات المتعلقة في الحصول على الوثائق الأمنية، وأن الغالبية العظمى من الطلبة (75%) يغطون نفقات دراستهم عن طريق العمل أو عن طريق عائلاتهم أو شخصيًا في ما كان هنالك جهات مختلفة مثل المنظمات الدولية، والجهات المانحة، ومنظمات أوروبية.

**منهج الوصفي-التحليلي:** استخدم المنهج الوصفي التحليلي لجمع البيانات، باستخدام استبانة تم توزيعها الكترونيا نظرًا لظروف هذه الجائحة، وتحليلها بهدف الإجابة عن أسئلة الدراسة، حيث يُعد هذا المنهج الأنسب لمثل هذه الدراسات. حيث تم طرح الأسئلة وبربطها بعلاقات مع اسئلة الدراسة، بهدف استخراج النتائج وفقًا لشواهد وقرائن متنوعة تم جمعها.

**مجتمع الدراسة:** تكوّن مجمع الدراسة من جميع الطلبة السوريين الدارسين في الجامعات الأردنية، والبالغ عددهم (15012) طالبًا وطالبة حسب إحصائيات وزارة التعليم العالي، خلال الفصل الدراسي الثاني للعام الجامعي 2018/2019م.

**عينة الدراسة:** تم أخذ عينة بالطريقة العشوائية البسيطة، تكونت من (257) طالبًا وطالبة، ويشكلون ما نسبته (1.71%) من مجتمع الدراسة، والجدول (1) يوضح توزيع أفراد العينة حسب متغيراتها.

**الجدول (1): توزيع أفراد عينة الدراسة حسب متغيراتها**

| المتغيرات | المستويات | التكرار | النسبة المئوية |
|---|---|---|---|
| الجنس | ذكر | 132 | 51.36% |
| | أنثى | 125 | 48.64% |
| الحالة الاجتماعية | أعزب | 231 | 89.88% |
| | غير ذلك | 26 | 10.12% |
| الفئة العمرية | أقل من 20 سنة | 109 | 42.41% |
| | من 20-24 سنة | 72 | 28.02% |
| | أكثر من 24 سنة | 76 | 29.57% |
| الإقليم | إقليم الشمال | 171 | 66.54% |
| | إقليم الوسط و إقليم الجنوب | 86 | 33.46% |
| مكان السكن | مدينة | 42 | 16.34% |
| | مخيم | 171 | 66.54% |
| | قرية | 44 | 17.12% |
| المستوى الدراسي | بكالوريوس | 231 | 89.88% |
| | دراسات عليا | 26 | 10.12% |
| المعدل التراكمي | ممتاز | 117 | 45.53% |
| | جيد جدًا | 67 | 26.07% |
| | جيد فأقل | 73 | 28.40% |
| صفة الجامعة | جامعة حكومية | 103 | 40.08% |
| | جامعة خاصة | 154 | 59.92% |
| هل تعمل | نعم | 40 | 15.56% |
| | لا | 217 | 84.44% |
| المجموع | | 257 | 100.00% |

154

**أداة الدراسة:** استخدم الباحثان استبانة "انعكاسات جائحة كورونا على الطلبة السوريين الدارسين في الجامعات الأردنية". تكونت الاستبانة من (39) فقرة موزعة على أربعة مجالات، وهي: المجال النفسي، وتضمن (14) فقرة، والمجال الأكاديمي، وتضمن (10) فقرات، والمجال الاقتصادي، وتضمن (7) فقرات، والمجال الاجتماعي، وتضمن (8) فقرات.

**صدق الاستبانة:** للتحقق من صدق الاستبانة، تم عرضها على لجنة من المحكمين والخبراء في الجامعات الأردنية، وعددهم (9) محكمين من ذوي الاختصاص والخبرة، وتم الأخذ بتوجيهات ومقترحات أعضاء اللجنة، فقد تم تعديل الصياغة اللغوية لبعض الفقرات عندما أجمع خمسة محكمين على ذلك.

**ثبات الاستبانة:** تم التحقق من ثبات الاستبانة، من خلال حساب معاملات الثبات لها باستخدام معادلة ألفا كرونباخ، حيث تم تطبيقها على عينة استطلاعية وعددهم (29) طالبًا وطالبة من خارج عينة الدراسة. تراوحت معاملات الثبات للمجالات بين (0.81 – 0.90)، وبلغت قيمة معامل الثبات للاستبانة الكلية (0.88). وهي قيم مقبولة لإجراء مثل هذه الدراسة (Lord, 1985).

**تصحيح الاستبانة:** تم استخدام تدريج ليكرت الخماسي لتقدير انعكاسات جائحة كورونا على الطلبة السوريين الدارسين في الجامعات الأردنية على النحو الآتي: (كبيرة جدًا، وكبيرة، ومتوسطة، وقليلة، وقليلة جدًا)، وتم إعطاء التقديرات الرقمية الآتية (5، 4، 3، 2، 1) على التوالي، وتم استخدام التدريج الإحصائي الآتي لتوزيع المتوسطات الحسابية (عودة، 2007):

- 1.00 - 2.49 انعكاسات بدرجة قليلة.
- 2.50 - 3.49 انعكاسات بدرجة متوسطة.
- 3.50 – 5.00 انعكاسات بدرجة كبيرة.

**متغيرات الدراسة:** اشتملت الدراسة على المتغيرات الآتية:

**أولًا: المتغيرات الوسيطة:**

**الجنس:** وله فئتان: (ذكور، وإناث).

**الحالة الاجتماعية:** وله فئتان: (أعزب، وغير ذلك).

**الفئة العمرية:** ولها ثلاث مستويات: (أقل من 20 سنة، ومن 20-24 سنة، وأكثر من 24 سنة).

**الإقليم:** وله فئتان: (إقليم الشمال، وإقليم الوسط وإقليم الجنوب).

**مكان السكن:** ولها ثلاث فئات: (مدينة، ومخيم، وقرية).

**المستوى الدراسي:** وله مستويان: (بكالوريوس، ودراسات عليا).

**المعدل التراكمي:** وله ثلاث مستويات: (ممتاز، وجيد جدًا، وجيد فأقل).

**صفة الجامعة:** وله فئتان: (جامعة حكومية، وجامعة خاصة).

**هل العمل:** وله فئتان: (نعم، لا).

**ثانيًا: المتغير التابع:** انعكاسات جائحة كورونا على الطلبة السوريين الدارسين في الجامعات الأردنية، التي يعبر عنها بالمتوسطات الحسابية لتقديرات أفراد العينة على فقرات ومجالات الاستبانة.

**المعالجات الإحصائية:** تم استخدام المتوسطات الحسابية والانحرافات المعيارية، واختبار تحليل التباين المتعدد، واختبار تحليل التباين التُساعي، واختبار شيفيه.

**عرض النتائج ومناقشتها:**

في ما يأتي عرضًا لنتائج الدراسة، بعد جمع البيانات بواسطة أداة الدراسة "استبانة انعكاسات جائحة كورونا على الطلبة السوريين الدارسين في الجامعات الأردنية"، وتم عرضها وفقًا لترتيب أسئلة الدراسة.

**عرض النتائج المتعلقة السؤال الأول ومناقشتها: "ما انعكاسات جائحة كورونا على الطلبة السوريين الدارسين في الجامعات الأردنية؟"**
للإجابة عن هذا السؤال، تم حساب المتوسطات الحسابية والانحرافات المعيارية لتقديرات أفراد العينة على مجالات الاستبانة، حيث كانت كما هي في الجدول (2).

155

الجدول (2): المتوسطات الحسابية والانحرافات المعيارية لتقديرات أفراد العينة على مجالات الاستبانة مرتبة تنازليًا حسب المتوسطات الحسابية

| الرتبة | الرقم | المجالات | المتوسط الحسابي* | الانحراف المعياري | درجة الانعكاسات |
|---|---|---|---|---|---|
| 1 | 3 | المجال الاقتصادي | 3.56 | 0.53 | كبيرة |
| 2 | 2 | المجال الأكاديمي | 3.50 | 0.36 | كبيرة |
| 3 | 4 | المجال الاجتماعي | 3.46 | 0.48 | متوسطة |
| 4 | 1 | المجال النفسي | 3.36 | 0.44 | متوسطة |
| | | الاستبانة ككل | 3.45 | 0.32 | كبيرة |

* الدرجة العظمى من (5)

يبين الجدول (2) أن "المجال الاقتصادي" قد احتل المرتبة الأولى بمتوسط حسابي (3.56) وانحراف معياري (0.53)، وجاء "المجال الاجتماعي" في المرتبة الثانية بمتوسط حسابي (3.50) وانحراف معياري (0.3.6)، وجاء "المجال النفسي" في المرتبة الأخيرة بمتوسط حسابي (3.36) وانحراف معياري (0.44)، وقد بلغ المتوسط الحسابي لتقديرات أفراد العينة على مجالات الاستبانة (3.45) بانحراف معياري (0.32)، وهو يقابل انعكاسات بدرجة متوسطة.

ويعزو الباحثان ذلك إلى أن المجال الاقتصادي هو الأشد تأثرًا، حيث تعطلت الحياة الاقتصادية ليس على مستوى الأردن فحسب، ولكن على مستوى العالم أجمع بسبب هذه الجائحة، فقد أدت عمليات الحظر سواء الجزئي أو الكلي إلى تعطل المنشآت والحياة الاقتصادية، وتأثر بذلك الدخول والمصروفات الأسرية التي تضاعفت من غير وجود مصادر للدخل.

أما بالنسبة للمجال الأكاديمي فقد تأثر نتيجة توجه الجامعات الى التعلم عن بُعد، الأمر الذي أحدث خللًا في العملية التعليمية التعلمية، وأدى إلى ارباك الطلبة، فهناك العديد من المساقات الجامعية التي لا يُجدي التعلم عن بُعد كثيرًا معها، وهناك مساقات بحاجة إلى النقاش والحوار والتدريب العملي فيها الأمر الذي لا توفره منصات التعلم الالكتروني.

وقد اتفقت هذه النتيجة مع نتائج دراسة الصندوق الائتماني الأوروبي (2018)، ونتائج دراسة ناصيف وكوموش (2020).

كما تم حساب المتوسطات الحسابية والانحرافات المعيارية لتقديرات أفراد العينة على فقرات مجالات الاستبانة، حيث كانت على النحو التالي:

المجال الأول: المجال النفسي:

الجدول (3): المتوسطات الحسابية والانحرافات المعيارية لتقديرات أفراد العينة على المجال النفسي مرتبة تنازليًا

| الرقم | الفقرات | المتوسط الحسابي* | الانحراف المعياري | درجة الانعكاسات |
|---|---|---|---|---|
| 6 | أعاني من تقلبات كبيرة في مزاجي | 3.93 | 1.08 | كبيرة |
| 10 | أعاني من سوء في حالتي النفسية بسبب ظروف الجائحة | 3.88 | 1.18 | كبيرة |
| 1 | ينتابني قلق شديد | 3.76 | 1.12 | كبيرة |
| 28 | اشعر بالوحدة لبعدي عن زملائي | 3.75 | 1.26 | كبيرة |
| 4 | اشعر بالحزن الشديد | 3.67 | 1.14 | كبيرة |
| 9 | يسيطر علي الخمول والكسل | 3.64 | 1.13 | كبيرة |
| 7 | أشعر براحة نفسية عند دخول الأماكن التي يتواجد فيها الناس | 3.28 | 1.11 | متوسطة |
| 5 | استطيع التكيف مع ظروف الجائحة المحيطة بي | 3.27 | 1.09 | متوسطة |
| 3 | افتقر إلى الشعور بالأمان | 3.26 | 1.22 | متوسطة |
| 2 | اشعر بالرعب الشديد من إصابتي بالمرض | 3.14 | 1.10 | متوسطة |
| 32 | أصبحت أسيرًا للألعاب الالكترونية | 3.05 | 1.39 | متوسطة |
| 17 | عززت جائحة كورونا ثقتي بقدراتي واعتمادي على نفسي في الدراسة | 2.83 | 1.24 | متوسطة |
| 38 | أشعر بالارتياح العام تجاه التعليمات الصادرة عن الجهات الرسمية | 2.77 | 1.31 | متوسطة |
| 8 | أشعر بالخوف عند خروجي من المنزل لقضاء حوائجي | 2.74 | 1.07 | متوسطة |
| | المجال ككل | 3.36 | 0.48 | متوسطة |

* الدرجة العظمى من (5)

156

يبين الجدول (3) أن الفقرة رقم (6) التي نصت على "أعاني من تقلبات كبيرة في مزاجي" احتلت المرتبة الأولى بمتوسط حسابي (3.93) وانحراف معياري (1.08)، وجاءت الفقرة رقم (10) التي كان نصها " أعاني من سوء في حالتي النفسية بسبب ظروف الجائحة" بالمرتبة الثانية بمتوسط حسابي (3.88) وانحراف معياري (1.18)، بينما احتلت الفقرة رقم (8) التي نصت على "أشعر بالخوف عند خروجي من المنزل لقضاء حوائجي" المرتبة الأخيرة بمتوسط حسابي (2.74) وانحراف معياري (1.07)، وقد بلغ المتوسط الحسابي لتقديرات أفراد العينة على هذا المجال ككل (3.46) وانحراف معياري (0.48)، وهو يقابل تقدير استخدام بدرجة كبيرة.

**المجال الثاني: المجال الأكاديمي:**

الجدول (4): المتوسطات الحسابية والانحرافات المعيارية لتقديرات أفراد العينة على المجال الأكاديمي مرتبة تنازليًا

| الرقم | الفقرات | المتوسط الحسابي* | الانحراف المعياري | درجة الانعكاسات |
|---|---|---|---|---|
| 11 | انظر بشوق للعودة إلى مقاعد الدراسة في الجامعة لأنها الطريقة الأفضل للتدريس | 4.37 | 1.08 | كبيرة |
| 13 | استمرار تفشي وباء كورونا يؤثر سلبًا في سير دراستي | 4.22 | 1.10 | كبيرة |
| 14 | التعلم عن بعد افقدني ركن أساسي في التعليم المتمثل في التفاعل والتواصل داخل المحاضرات | 4.19 | 1.30 | كبيرة |
| 12 | عملية التعلم عن بعد أضعفت روح المنافسة عندي | 4.10 | 1.28 | كبيرة |
| 29 | افتقد النشاطات اللامنهجية التي كنت أمارسها في الجامعة | 4.10 | 1.07 | كبيرة |
| 35 | تشكلت لدى طرق جديدة للتفكير خلال جائحة كورونا | 3.53 | 1.22 | كبيرة |
| 19 | تابعت دروسي من خلال التعلم عن بعد بطريقة سهلة وميسرة | 2.73 | 1.21 | متوسطة |
| 16 | استراتيجية التعلم عن بعد أسهمت في منحي وقتًا أكبر للدراسة | 2.72 | 1.24 | متوسطة |
| 18 | ساهمت وسائل التكنولوجيا الحديثة إيجابًا في رفع تحصيلي الأكاديمي | 2.66 | 1.25 | متوسطة |
| 15 | تجربتي في التعلم عن بعد ناجحة وأرغب بالاستمرار حتى بعد انتهاء جائحة كورونا | 2.37 | 1.33 | قليلة |
| المجال ككل | | 3.50 | 0.36 | كبيرة |

* الدرجة العظمى من (5)

يبين الجدول (4) أن الفقرة رقم (11) التي نصت على "انظر بشوق للعودة إلى مقاعد الدراسة في الجامعة لأنها الطريقة الأفضل للتدريس" قد احتلت المرتبة الأولى بمتوسط حسابي (4.37) وانحراف معياري (1.08)، وجاءت الفقرة رقم (13) التي كان نصها "استمرار تفشي وباء كورونا يؤثر سلبًا في سير دراستي" بالمرتبة الثانية بمتوسط حسابي (4.22) وانحراف معياري (1.10)، بينما احتلت الفقرة رقم (15) التي نصت على "تجربتي في التعلم عن بعد ناجحة وأرغب بالاستمرار حتى بعد انتهاء جائحة كورونا" المرتبة الأخيرة بمتوسط حسابي (2.37) وانحراف معياري (1.33)، وقد بلغ المتوسط الحسابي لتقديرات أفراد العينة على هذا المجال ككل (3.50) وانحراف معياري (0.62)، وهو يقابل تقدير انعكاسات بدرجة كبيرة.

**المجال الثالث: المجال الاقتصادي:**

الجدول (5): المتوسطات الحسابية والانحرافات المعيارية لتقديرات أفراد العينة على المجال الاقتصادي مرتبة تنازليًا

| الرقم | الفقرات | المتوسط الحسابي* | الانحراف المعياري | درجة الانعكاسات |
|---|---|---|---|---|
| 23 | تواجه عائلتي أزمة اقتصادية بسبب المستلزمات الإضافية لجائحة كورونا | 4.19 | 1.08 | كبيرة |
| 21 | ابحث عن فرصة عمل للإنفاق على نفسي | 4.13 | 1.26 | كبيرة |
| 24 | انخفض دخل عائلتي خلال جائحة كورونا | 4.11 | 1.18 | كبيرة |
| 26 | تعطل بعض أفراد أسرتي عن العمل بسبب جائحة كورونا | 3.96 | 1.16 | كبيرة |
| 22 | سأضطر لتأجيل دراستي بسبب ظروف الاقتصادية خلال الجائحة. | 3.45 | 1.27 | متوسطة |
| 20 | أسهمت جائحة كرونا في تقليل الأعباء المالية على الطلبة وذويهم من خلال نهج استراتيجية التعلم عن بعد | 2.90 | 1.43 | متوسطة |
| 25 | الوضع الاقتصادي لعائلتي خلال جائحة كورونا مستقر ولا تأثير عليه | 2.22 | 1.23 | قليلة |
| المجال ككل | | 3.56 | 0.53 | كبيرة |

* الدرجة العظمى من (5)

يبين الجدول (5) أن الفقرة رقم (23) التي نصت على "تواجه عائلتي أزمة اقتصادية بسبب المستلزمات الإضافية لجائحة كورونا" قد احتلت المرتبة الأولى بمتوسط حسابي (4.19) وانحراف معياري (1.08)، وجاءت الفقرة رقم (21) التي كان نصها "ابحث عن فرصة عمل للإنفاق على نفسي" المرتبة الثانية بمتوسط حسابي (4.13) وانحراف معياري (1.26)، بينما احتلت الفقرة رقم (25) التي نصت على "الوضع الاقتصادي لعائلتي خلال جائحة كورونا مستقر ولا تأثير عليه" المرتبة الأخيرة بمتوسط حسابي (2.22) وانحراف معياري (1.23)، وقد بلغ المتوسط الحسابي لتقديرات أفراد العينة على هذا المجال ككل (3.56) وانحراف معياري (0.53)، وهو يقابل انعكاسات بدرجة كبيرة.

**المجال الرابع: المجال الاجتماعي:**

**الجدول (6): المتوسطات الحسابية والانحرافات المعيارية لتقديرات أفراد العينة على المجال الاجتماعي مرتبة تنازليًا**

| الرقم | الفقرات | المتوسط الحسابي* | الانحراف المعياري | درجة الانعكاسات |
|---|---|---|---|---|
| 36 | شعرت بقيمة الأصدقاء خلال فترة الحجر | 3.77 | 1.17 | كبيرة |
| 39 | حرصي على سلامة عائلتي دفعني إلى الالتزام بالنزل وعدم الخروج إلا للضرورة | 3.75 | 1.18 | كبيرة |
| 31 | افتقدت العلاقات الاجتماعية المباشرة مع أقاربي | 3.66 | 1.21 | كبيرة |
| 30 | أصبحت علاقاتي مع أفراد عائلتي متوترة | 3.47 | 1.22 | متوسطة |
| 37 | كثرت المشاجرات والنزاعات في منطقتنا | 3.36 | 1.26 | متوسطة |
| 34 | أسهمت جائحة كورونا في بث روح المحبة والألفة بين أفراد عائلتي | 3.25 | 1.35 | متوسطة |
| 27 | استمرت علاقاتي الاجتماعية خلال أزمة كورونا ولم انقطع عن الأقارب والأصدقاء من خلال استخدامي لمواقع التواصل الاجتماعي | 3.21 | 1.30 | متوسطة |
| 33 | اكتسبت قيمًا اجتماعية جديدة خلال هذه الجائحة | 3.17 | 1.18 | متوسطة |
| | المجال ككل | 3.46 | 0.48 | متوسطة |

* الدرجة العظمى من (5)

يبين الجدول (6) أن الفقرة رقم (36) التي نصت على "شعرت بقيمة الأصدقاء خلال فترة الحجر" قد احتلت المرتبة الأولى بمتوسط حسابي (3.77) وانحراف معياري (1.17)، وجاءت الفقرة رقم (39) التي كان نصها "حرصي على سلامة عائلتي دفعني إلى الالتزام بالنزل وعدم الخروج إلا للضرورة" المرتبة الثانية بمتوسط حسابي (3.75) وانحراف معياري (1.18)، بينما احتلت الفقرة رقم (33) التي نصت على "اكتسبت قيمًا اجتماعية جديدة خلال هذه الجائحة" المرتبة الأخيرة بمتوسط حسابي (3.17) وانحراف معياري (1.18)، وقد بلغ المتوسط الحسابي لتقديرات أفراد العينة على هذا المجال ككل (3.46) وانحراف معياري (0.48)، وهو يقابل انعكاسات بدرجة متوسطة.

**النتائج المتعلقة بالسؤال الثاني ومناقشتها:** "هل هناك فروق ذات دلالة احصائية عند مستوى الدلالة ( α≤0.05) بين تقديرات أفراد العينة لانعكاسات جائحة كورونا على الطلبة السوريين الدارسين في الجامعات الأردنية تعزى لمتغيرات الجنس، والحالة الاجتماعية، والفئة العمرية، والإقليم، ومكان السكن، والمستوى الدراسي، والمعدل التراكمي، وصفة الجامعة، وهل تعمل؟"

للإجابة عن هذا السؤال، تم حساب المتوسطات الحسابية والانحرافات المعيارية لتقديرات أفراد العينة لانعكاسات جائحة كورونا على الطلبة السوريين الدارسين في الجامعات الأردنية على مجالات أداة الدراسة والأداة الكلية حسب متغيراتها، حيث كانت على النحو الآتي:

**أ- حسب متغير الجنس:**

**الجدول (7): المتوسطات الحسابية والانحرافات المعيارية لتقديرات أفراد عينة الدراسة لانعكاسات جائحة كورونا على الطلبة السوريين الدارسين في الجامعات الأردنية حسب متغير الجنس**

| المجالات | ذكر (ن = 132) | | أنثى (ن = 125) | |
|---|---|---|---|---|
| | المتوسط الحسابي | الانحراف المعياري | المتوسط الحسابي | الانحراف المعياري |
| المجال النفسي | 3.40 | .450 | 3.31 | .416 |
| المجال الأكاديمي | 3.49 | .382 | 3.51 | .326 |
| المجال الاقتصادي | 3.60 | .527 | 3.53 | .538 |
| المجال الاجتماعي | 3.49 | .519 | 3.42 | .426 |
| الاستبانة ككل | 3.48 | .351 | 3.42 | .276 |

158

ب- حسب متغير الحالة الاجتماعية:

الجدول (8): المتوسطات الحسابية والانحرافات المعيارية لتقديرات أفراد عينة الدراسة لانعكاسات جائحة كورونا على الطلبة السوريين الدارسين في الجامعات الأردنية حسب متغير الحالة الاجتماعية

| المجالات | أعزب (ن = 231) | | غير ذلك (ن = 26) | |
|---|---|---|---|---|
| | المتوسط الحسابي | الانحراف المعياري | المتوسط الحسابي | الانحراف المعياري |
| المجال النفسي | 3.38 | .427 | 3.10 | .439 |
| المجال الأكاديمي | 3.51 | .363 | 3.43 | .277 |
| المجال الاقتصادي | 3.57 | .530 | 3.51 | .557 |
| المجال الاجتماعي | 3.46 | .481 | 3.42 | .446 |
| الاستبانة ككل | 3.46 | .318 | 3.33 | .292 |

ج- حسب متغير الفئة العمرية:

الجدول (9): المتوسطات الحسابية والانحرافات المعيارية لتقديرات أفراد عينة الدراسة لانعكاسات جائحة كورونا على الطلبة السوريين الدارسين في الجامعات الأردنية حسب متغير الفئة العمرية

| المجالات | أقل من 20 سنة (ن = 109) | | من 20-24 سنة (ن = 72) | | أكثر من 24 سنة (ن = 76) | |
|---|---|---|---|---|---|---|
| | المتوسط الحسابي | الانحراف المعياري | المتوسط الحسابي | الانحراف المعياري | المتوسط الحسابي | الانحراف المعياري |
| المجال النفسي | 3.33 | .419 | 3.23 | .528 | 3.44 | .452 |
| المجال الأكاديمي | 3.49 | .335 | 3.44 | .264 | 3.52 | .410 |
| المجال الاقتصادي | 3.46 | .538 | 3.80 | .428 | 3.76 | .466 |
| المجال الاجتماعي | 3.39 | .466 | 3.45 | .583 | 3.60 | .459 |
| الاستبانة ككل | 3.41 | .308 | 3.43 | .351 | 3.55 | .315 |

د- حسب متغير الإقليم:

الجدول (10): المتوسطات الحسابية والانحرافات المعيارية لتقديرات أفراد عينة الدراسة لانعكاسات جائحة كورونا على الطلبة السوريين الدارسين في الجامعات الأردنية حسب متغير الإقليم

| المجالات | إقليم الشمال (ن = 171) | | إقليم الوسط والجنوب (ن = 86) | |
|---|---|---|---|---|
| | المتوسط الحسابي | الانحراف المعياري | المتوسط الحسابي | الانحراف المعياري |
| المجال النفسي | 3.35 | .453 | 3.36 | .403 |
| المجال الأكاديمي | 3.50 | .368 | 3.50 | .331 |
| المجال الاقتصادي | 3.56 | .583 | 3.58 | .415 |
| المجال الاجتماعي | 3.49 | .515 | 3.39 | .381 |
| الاستبانة ككل | 3.45 | .339 | 3.44 | .272 |

هـ- حسب متغير مكان السكن:

الجدول (11): المتوسطات الحسابية والانحرافات المعيارية لتقديرات أفراد عينة الدراسة لانعكاسات جائحة كورونا على الطلبة السوريين الدارسين في الجامعات الأردنية حسب متغير مكان السكن

| المجالات | مدينة (ن = 42) | | مخيم (ن = 171) | | قرية (ن = 44) | |
|---|---|---|---|---|---|---|
| | المتوسط الحسابي | الانحراف المعياري | المتوسط الحسابي | الانحراف المعياري | المتوسط الحسابي | الانحراف المعياري |
| المجال النفسي | 3.29 | .466 | 3.37 | .430 | 3.37 | .436 |
| المجال الأكاديمي | 3.47 | .456 | 3.50 | .341 | 3.46 | .304 |
| المجال الاقتصادي | 3.46 | .530 | 3.58 | .547 | 3.61 | .473 |
| المجال الاجتماعي | 3.42 | .423 | 3.45 | .484 | 3.48 | .502 |
| الاستبانة ككل | 3.35 | .319 | 3.45 | .329 | 3.46 | .276 |

و- حسب متغير المستوى الدراسي:

الجدول (12): المتوسطات الحسابية والانحرافات المعيارية لتقديرات أفراد عينة الدراسة لانعكاسات جائحة كورونا على الطلبة السوريين الدارسين في الجامعات الأردنية حسب متغير المستوى الدراسي

| المجالات | بكالوريوس (ن = 231) | | دراسات عليا (ن = 26) | |
|---|---|---|---|---|
| | المتوسط الحسابي | الانحراف المعياري | المتوسط الحسابي | الانحراف المعياري |
| المجال النفسي | 3.37 | .439 | 3.21 | .392 |
| المجال الأكاديمي | 3.49 | .345 | 3.60 | .432 |
| المجال الاقتصادي | 3.55 | .533 | 3.71 | .509 |
| المجال الاجتماعي | 3.45 | .453 | 3.53 | .652 |
| الاستبانة ككل | 3.45 | .310 | 3.47 | .384 |

ز- حسب متغير المعدل التراكمي:

الجدول (13): المتوسطات الحسابية والانحرافات المعيارية لتقديرات أفراد عينة الدراسة لانعكاسات جائحة كورونا على الطلبة السوريين الدارسين في الجامعات الأردنية حسب متغير المعدل التراكمي

| المجالات | ممتاز (ن = 117) | | جيد جدًا (ن = 67) | | جيد فأقل (ن = 73) | |
|---|---|---|---|---|---|---|
| | المتوسط الحسابي | الانحراف المعياري | المتوسط الحسابي | الانحراف المعياري | المتوسط الحسابي | الانحراف المعياري |
| المجال النفسي | 3.25 | .389 | 3.21 | .470 | 3.42 | .450 |
| المجال الأكاديمي | 3.46 | .298 | 3.48 | .370 | 3.59 | .412 |
| المجال الاقتصادي | 3.51 | .550 | 3.61 | .516 | 3.60 | .517 |
| المجال الاجتماعي | 3.49 | .435 | 3.37 | .467 | 3.48 | .541 |
| الاستبانة ككل | 3.40 | .290 | 3.38 | .323 | 3.51 | .346 |

ح- حسب متغير صفة الجامعة:

الجدول (14): المتوسطات الحسابية والانحرافات المعيارية لتقديرات أفراد عينة الدراسة لانعكاسات جائحة كورونا على الطلبة السوريين الدارسين في الجامعات الأردنية حسب متغير صفة الجامعة

| المجالات | جامعة حكومية (ن = 103) | | جامعة خاصة (ن = 154) | |
|---|---|---|---|---|
| | المتوسط الحسابي | الانحراف المعياري | المتوسط الحسابي | الانحراف المعياري |
| المجال النفسي | 3.25 | .450 | 3.42 | .414 |
| المجال الأكاديمي | 3.47 | .381 | 3.52 | .338 |
| المجال الاقتصادي | 3.42 | .645 | 3.66 | .415 |
| المجال الاجتماعي | 3.40 | .545 | 3.50 | .421 |
| الاستبانة ككل | 3.37 | .341 | 3.50 | .289 |

ي- حسب متغير هل تعمل:

الجدول (15): المتوسطات الحسابية والانحرافات المعيارية لتقديرات أفراد عينة الدراسة لانعكاسات جائحة كورونا على الطلبة السوريين الدارسين في الجامعات الأردنية حسب متغير هل تعمل

| المجالات | نعم (ن = 40) | | لا (ن = 217) | |
|---|---|---|---|---|
| | المتوسط الحسابي | الانحراف المعياري | المتوسط الحسابي | الانحراف المعياري |
| المجال النفسي | 3.35 | .548 | 3.36 | .413 |
| المجال الأكاديمي | 3.50 | .374 | 3.50 | .353 |
| المجال الاقتصادي | 3.74 | .427 | 3.53 | .544 |
| المجال الاجتماعي | 3.41 | .650 | 3.46 | .439 |
| الاستبانة ككل | 3.47 | .429 | 3.45 | .294 |

يتبين من الجداول (7، 8، 9، 10، 11، 12، 13، 14، 15) وجود فروق ظاهرية بين المتوسطات الحسابية لتقديرات أفراد عينة الدراسة لانعكاسات جائحة كورونا على الطلبة السوريين الدارسين في الجامعات الأردنية حسب متغيرات الدراسة، ولتعرُّف مستويات الدلالة الإحصائية لتلك الفروق تم استخدام تحليل التباين المتعدد، والجدول (16) يبين ذلك.

**الجدول (16): نتائج تحليل التباين المتعدد للفروق بين المتوسطات الحسابية لتقديرات أفراد العينة لانعكاسات جائحة كورونا على الطلبة السوريين الدارسين في الجامعات الأردنية حسب متغيرات الدراسة**

| مصدر التباين | المجالات | مجموع المربعات | درجات الحرية | متوسط المربعات | قيمة ف | الدلالة الإحصائية |
|---|---|---|---|---|---|---|
| الجنس<br>قيمة هوتلنغ= 0.016<br>ح= 0.548 | المجال النفسي | .124 | 1 | .124 | .741 | .390 |
| | المجال الأكاديمي | .016 | 1 | .016 | .130 | .719 |
| | المجال الاقتصادي | .131 | 1 | .131 | .537 | .464 |
| | المجال الاجتماعي | .143 | 1 | .143 | .652 | .420 |
| الحالة الاجتماعية<br>قيمة هوتلنغ = 0.064<br>ح= 0.005 | المجال النفسي | 2.179 | 1 | 2.179 | 13.025 | .000* |
| | المجال الأكاديمي | .097 | 1 | .097 | .782 | .377 |
| | المجال الاقتصادي | .201 | 1 | .201 | .822 | .365 |
| | المجال الاجتماعي | .015 | 1 | .015 | .067 | .796 |
| الفئة العمرية<br>قيمة ولكس= 0.946<br>ح= 0.099 | المجال النفسي | .974 | 2 | .487 | 2.910 | .056 |
| | المجال الأكاديمي | .266 | 2 | .133 | 1.073 | .344 |
| | المجال الاقتصادي | .477 | 2 | .239 | .978 | .377 |
| | المجال الاجتماعي | .178 | 2 | .089 | .408 | .665 |
| الإقليم<br>قيمة هوتلنغ= 0.031<br>ح= 0.504 | المجال النفسي | .167 | 1 | .167 | 1.000 | .318 |
| | المجال الأكاديمي | .002 | 1 | .002 | .016 | .899 |
| | المجال الاقتصادي | .004 | 1 | .004 | .017 | .895 |
| | المجال الاجتماعي | .515 | 1 | .515 | 2.357 | .126 |
| مكان السكن<br>قيمة ولكس= 0.989<br>ح= 0.001 | المجال النفسي | .204 | 2 | .102 | .609 | .545 |
| | المجال الأكاديمي | .024 | 2 | .012 | .098 | .907 |
| | المجال الاقتصادي | 5.783 | 2 | 2.891 | 11.850 | .000* |
| | المجال الاجتماعي | .958 | 2 | .479 | 2.191 | .114 |
| المستوى الدراسي<br>قيمة هوتلنغ= 0.047<br>ح= 0.026 | المجال النفسي | .556 | 1 | .556 | 3.325 | .069 |
| | المجال الأكاديمي | .450 | 1 | .450 | 3.630 | .058 |
| | المجال الاقتصادي | .233 | 1 | .233 | .953 | .330 |
| | المجال الاجتماعي | .094 | 1 | .094 | .431 | .512 |
| المعدل التراكمي<br>قيمة ولكس= 0.903<br>ح= 0.002 | المجال النفسي | 1.357 | 2 | .678 | 4.054 | .019* |
| | المجال الأكاديمي | 1.392 | 2 | .696 | 5.613 | .004* |
| | المجال الاقتصادي | .907 | 2 | .454 | 1.859 | .158 |
| | المجال الاجتماعي | 1.091 | 2 | .546 | 2.497 | .084 |
| صفة الجامعة<br>قيمة هوتلنغ= 0.072<br>ح= 0.002 | المجال النفسي | 1.737 | 1 | 1.737 | 10.384 | .001* |
| | المجال الأكاديمي | .323 | 1 | .323 | 2.608 | .108 |
| | المجال الاقتصادي | 3.025 | 1 | 3.025 | 12.397 | .001* |
| | المجال الاجتماعي | .731 | 1 | .731 | 3.344 | .069 |
| هل تعمل<br>قيمة هوتلنغ= 0.066<br>ح= 0.004 | المجال النفسي | .544 | 1 | .544 | 3.251 | .073 |
| | المجال الأكاديمي | .007 | 1 | .007 | .057 | .811 |
| | المجال الاقتصادي | 1.376 | 1 | 1.376 | 5.639 | .018* |
| | المجال الاجتماعي | .423 | 1 | .423 | 1.937 | .165 |

161

| مصدر التباين | المجالات | مجموع المربعات | درجات الحرية | متوسط المربعات | قيمة ف |
|---|---|---|---|---|---|
| الخطأ | المجال النفسي | 40.826 | 244 | .167 | |
| | المجال الأكاديمي | 30.264 | 244 | .124 | |
| | المجال الاقتصادي | 59.532 | 244 | .244 | |
| | المجال الاجتماعي | 53.326 | 244 | .219 | |

● ذات دلالة إحصائية عند مستوى الدلالة ($\alpha \geq 0.05$).

يبين الجدول (16):

1- عدم وجود فروق ذات دلالة إحصائية بين متوسط تقديرات أفراد عينة الدراسة عند جميع مجالات انعكاسات جائحة كورونا على الطلبة السوريين الدارسين في الجامعات الأردنية تعزى لمتغير الجنس.

ويمكن عزو هذه النتيجة إلى أن الجميع قد تأثر في هذه الجائحة سواء الذكور أم الاناث.

2- عدم وجود فروق ذات دلالة إحصائية بين متوسط تقديرات أفراد عينة الدراسة عند جميع مجالات انعكاسات جائحة كورونا على الطلبة السوريين الدارسين في الجامعات الأردنية تعزى لمتغير الحالة الاجتماعية، باستثناء المجال النفسي، وذلك لصالح تقديرات (أعزب).

وقد يُعزى ذلك إلى أن الطلبة ذوي الحالة الاجتماعية (أعزب)، ونتيجة كانت آثارها النفسية أشد لغيات العلاقات الأسرية والاجتماعية.

3- عدم وجود فروق ذات دلالة إحصائية بين متوسط تقديرات أفراد عينة الدراسة عند جميع مجالات انعكاسات جائحة كورونا على الطلبة السوريين الدارسين في الجامعات الأردنية تعزى لمتغير الفئة العمرية.

ويمكن أن يرجع السبب في ذلك إلى أن انعكاسات جائحة كورونا طالت جميع فئات المجتمع بغض النظر عن فئاتهم العمرية.

4- عدم وجود فروق ذات دلالة إحصائية بين متوسط تقديرات أفراد عينة الدراسة عند جميع مجالات انعكاسات جائحة كورونا على الطلبة السوريين الدارسين في الجامعات الأردنية تعزى لمتغير الإقليم.

ويمكن تفسير ذلك أن أيام الحظر والاغلاقات كانت تشمل جميع أقاليم المملكة دون استثناء، وكانت الاجراءات الحكومية من تعطيل القطاعات الاقتصادية تشمل جميع أنحاء عدم وجود فروق ذات دلالة إحصائية بين متوسط تقديرات أفراد عينة الدراسة عند جميع مجالات انعكاسات جائحة كورونا على الطلبة السوريين الدارسين في الجامعات الأردنية، باستثناء المجال الاقتصادي تعزى لمتغير مكان السكن. ولتحديد مصادر تلك الفروق تم استخدام اختبار شافيه ('Scheffe) كما هو موضح في الجدول (17).

**الجدول (17): نتائج اختبار شافيه ('Scheffe) للفروق بين تقديرات أفراد العينة على المجال الاقتصادي حسب متغير مكان السكن**

| مكان السكن | | مدينة | مخيم | قرية |
|---|---|---|---|---|
| | المتوسط الحسابي | 3.46 | 3.58 | 3.61 |
| مدينة | 3.46 | | *0.12 | *0.15 |
| مخيم | 3.58 | | | 0.03 |
| قرية | 3.61 | | | |

● ذات دلالة إحصائية عند مستوى الدلالة ($\alpha \geq 0.05$).

يبين الجدول (17) أن هناك فروقًا ذات دلالة إحصائية بين متوسط تقديرات ذوي مكان السكن (مدينة) من جهة، ومتوسط تقديرات ذوي مكان السكن (مخيم، وقرية) من جهة ثانية، تعزى لمتغير مكان السكن، وذلك لصالح تقديرات ذوي مكان السكن (مخيم، وقرية).

ويمكن تفسير ذلك إلى طبيعة السكن في القرية أو المخيم تكون عادة تكاليف المعيشة الاقتصادية فيهما أقل من تكاليف المعيشة في المدن.

5- عدم وجود فروق ذات دلالة إحصائية بين متوسط تقديرات أفراد عينة الدراسة عند جميع مجالات انعكاسات جائحة كورونا على الطلبة السوريين الدارسين في الجامعات الأردنية تعزى لمتغير المستوى الدراسي.

6- عدم وجود فروق ذات دلالة إحصائية بين متوسط تقديرات أفراد عينة الدراسة عند مجالي (المجال الاقتصادي، والمجال الاجتماعي) من مجالات انعكاسات جائحة كورونا على الطلبة السوريين الدارسين في الجامعات الأردنية، وعدم وجود فروق ذات دلالة إحصائية بين متوسط تقديرات أفراد عينة الدراسة عند مجالي (المجال النفسي، والمجال الأكاديمي) تعزى لمتغير المعدل التراكمي. ولتحديد مصادر تلك الفروق تم استخدام اختبار شافيه ('Scheffe) كما هو موضح في الجدول (18).

**الجدول (18):** نتائج اختبار شافيه ('Scheffe) للفروق بين تقديرات أفراد العينة على المجال النفسي والمجال الأكاديمي حسب متغير المعدل التراكمي

| المجال | المعدل التراكمي | | ممتاز | جيد جدًا | جيد فأقل |
|---|---|---|---|---|---|
| | المتوسط الحسابي | | 3.25 | 3.21 | 3.42 |
| المجال النفسي | ممتاز | 3.25 | | 0.04 | 0.17* |
| | جيد جدًا | 3.21 | | | 0.21* |
| | جيد فأقل | 3.42 | | | |
| المجال | المعدل التراكمي | | 3.46 | 3.48 | 3.59 |
| | المتوسط الحسابي | | | | |
| المجال الأكاديمي | ممتاز | 3.46 | | 0.02 | 0.13* |
| | جيد جدًا | 3.48 | | | 0.11* |
| | جيد فأقل | 3.59 | | | |

- ذات دلالة إحصائية عند مستوى الدلالة (α≥0.05).

يبين الجدول (18) أن هناك فروقًا ذات دلالة إحصائية بين متوسط تقديرات ذوي المعدل التراكمي (جيد فأقل) من جهة، ومتوسط تقديرات ذوي المعدل التراكمي (ممتاز، وجيد جدًا) من جهة ثانية، تعزى لمتغير المعدل التراكمي، وذلك لصالح تقديرات ذوي المعدل التراكمي (جيد فأقل).

7- عدم وجود فروق ذات دلالة إحصائية بين متوسط تقديرات أفراد عينة الدراسة عند جميع مجالات انعكاسات جائحة كورونا على الطلبة السوريين الدارسين في الجامعات الأردنية تعزى لمتغير صفة الجامعة، باستثناء المجال النفسي، والمجال الاقتصادي، وذلك لصالح تقديرات ذوي نوع الجامعة (جامعة خاصة).

ويمكن عزو هذه النتيجة إلى أن الجامعات الخاصة تتطلب أقساطًا دراسية أعلى من الجامعات الحكومية.
ولم يعثر الباحثان على أية دراسة تناولت هذا المتغير

8- عدم وجود فروق ذات دلالة إحصائية بين متوسط تقديرات أفراد عينة الدراسة عند جميع مجالات انعكاسات جائحة كورونا على الطلبة السوريين الدارسين في الجامعات الأردنية تعزى لمتغير العمل، باستثناء المجال الاقتصادي، وذلك لصالح تقديرات (نعم أعمل).

ويمكن أن يرجع السبب في ذلك إلى أن الفئة التي كانت تعمل، كانت تُغطي جزءًا من تلك النفقات من خلال كسبها من العمل، وعند قدوم الجائحة انقطعت تلك الإيرادات من خلال العمل.

وقد اتفقت هذه النتيجة مع نتائج دراسة الصندوق الائتماني الأوروبي (2018).

كما تم إجراء اختبار تحليل التباين التُساعي للفروق بين تقديرات أفراد العينة على مجالات انعكاسات جائحة كورونا على الطلبة السوريين الدارسين في الجامعات الأردنية ككل حسب متغيرات الدراسة، حيث كانت النتائج كما هي موضحة في الجدول (19).

**الجدول (19):** اختبار تحليل التباين التُساعي للفروق بين تقديرات أفراد العينة على مجالات انعكاسات جائحة كورونا على الطلبة السوريين الدارسين في الجامعات الأردنية ككل حسب متغيرات الدراسة

| المتغيرات | مجموع المربعات | درجة الحرية | متوسط المربعات | قيمة ف | الدلالة الإحصائية |
|---|---|---|---|---|---|
| الجنس | .011 | 1 | .011 | .124 | .725 |
| الحالة الاجتماعية | .511 | 1 | .511 | 5.614 | .019* |
| الفئة العمرية | .116 | 2 | .058 | .635 | .531 |
| الإقليم | .086 | 1 | .086 | .948 | .331 |
| مكان السكن | .618 | 2 | .309 | 3.392 | .035* |
| المستوى الدراسي | .003 | 1 | .003 | .032 | .859 |
| المعدل التراكمي | .879 | 2 | .439 | 4.825 | .009* |
| صفة الجامعة | 1.224 | 1 | 1.224 | 13.445 | .000* |
| هل تعمل | .028 | 1 | .028 | .303 | .583 |
| الخطأ | 22.220 | 244 | .091 | | |
| الكلي | 3085.072 | 256 | | | |

- ذات دلالة إحصائية عند مستوى الدلالة (α≥0.05).

163

يبين الجدول (19):

1- عدم وجود فروق ذات دلالة إحصائية بين متوسط تقديرات أفراد عينة الدراسة عند جميع مجالات انعكاسات جائحة كورونا على الطلبة السوريين الدارسين في الجامعات الأردنية تعزى لمتغيرات الجنس، الفئة العمرية، والإقليم، والمستوى الدراسي، والعمل.

2- وجود فروق ذات دلالة إحصائية بين متوسط تقديرات أفراد عينة الدراسة عند جميع مجالات انعكاسات جائحة كورونا على الطلبة السوريين الدارسين في الجامعات الأردنية ككل تعزى لمتغير الحالة الاجتماعية، وذلك لصالح تقديرات (أعزب).

3- وجود فروق ذات دلالة إحصائية بين متوسط تقديرات أفراد عينة الدراسة عند جميع مجالات انعكاسات جائحة كورونا على الطلبة السوريين الدارسين في الجامعات الأردنية ككل، تعزى لمتغير مكان السكن. ولتحديد مصادر تلك الفروق تم استخدام اختبار شافيه ('Scheffe) كما هو موضح في الجدول (20).

**الجدول (20):** نتائج اختبار شافيه ('Scheffe) للفروق بين تقديرات أفراد العينة على جميع مجالات انعكاسات جائحة كورونا على الطلبة السوريين الدارسين في الجامعات الأردنية ككل حسب متغير مكان السكن

| مكان السكن | | مدينة | مخيم | قرية |
|---|---|---|---|---|
| المتوسط الحسابي | | 3.35 | 3.45 | 3.46 |
| مدينة | 3.35 | | *0.10 | *0.11 |
| مخيم | 3.45 | | | 0.01 |
| قرية | 3.46 | | | |

● ذات دلالة إحصائية عند مستوى الدلالة ($\alpha \geq 0.05$).

يبين الجدول (20) أن هناك فروقًا ذات دلالة إحصائية بين متوسط تقديرات ذوي مكان السكن (مدينة) من جهة، ومتوسط تقديرات ذوي مكان السكن (مخيم، وقرية) من جهة ثانية، تعزى لمتغير مكان السكن، وذلك لصالح تقديرات ذوي مكان السكن (مخيم، وقرية).

4- وجود فروق ذات دلالة إحصائية بين متوسط تقديرات أفراد عينة الدراسة عند جميع مجالات انعكاسات جائحة كورونا على الطلبة السوريين الدارسين في الجامعات الأردنية ككل، تعزى لمتغير المعدل التراكمي. ولتحديد مصادر تلك الفروق تم استخدام اختبار شافيه ('Scheffe) كما هو موضح في الجدول (21).

**الجدول (21):** نتائج اختبار شافيه ('Scheffe) للفروق بين تقديرات أفراد العينة عند جميع مجالات انعكاسات جائحة كورونا على الطلبة السوريين الدارسين في الجامعات الأردنية ككل حسب متغير المعدل التراكمي

| المعدل التراكمي | | ممتاز | جيد جدًا | جيد فأقل |
|---|---|---|---|---|
| المتوسط الحسابي | | 3.40 | 3.38 | 3.51 |
| ممتاز | 3.40 | | 0.02 | *0.11 |
| جيد جدًا | 3.38 | | | *0.13 |
| جيد فأقل | 3.51 | | | |

● ذات دلالة إحصائية عند مستوى الدلالة ($\alpha \geq 0.05$).

يبين الجدول (18) أن هناك فروقًا ذات دلالة إحصائية بين متوسط تقديرات ذوي المعدل التراكمي (جيد فأقل) من جهة، ومتوسط تقديرات ذوي المعدل التراكمي (ممتاز، وجيد جدًا) من جهة ثانية، تعزى لمتغير المعدل التراكمي، وذلك لصالح تقديرات ذوي المعدل التراكمي (جيد فأقل).

5- وجود فروق ذات دلالة إحصائية بين متوسط تقديرات أفراد عينة الدراسة عند جميع مجالات انعكاسات جائحة كورونا على الطلبة السوريين الدارسين في الجامعات الأردنية ككل تعزى لمتغير صفة الجامعة، وذلك لصالح تقديرات ذوي نوع الجامعة (جامعة خاصة).

**خاتمة:**

فقد استهدفت الدراسة قراءة لواقع أثر جائحة كورونا بموجتيها الأولى والثانية في الطلبة اللاجئين بالجامعات الأردنية التي انبثق عنها عدد من الآثار الجانبية التي انعكست عليهم وعلى هذا لقد انجلت الدراسة عن عدد من التوصيات والنتائج:

● تنفيذ برامج إرشادية ونفسية لمساعدة الطلبة اللاجئين السوريين الدارسين في الجامعات الأردنية في تخطي الآثار النفسية للجائحة.

- تنفيذ برامج دعم اقتصادية لمساعدة الطلبة اللاجئين السوريين الدارسين في الجامعات الأردنية في التغلب على المشكلات المالية التي نتجت عن هذه الجائحة.

- إجراء مزيدًا من الدراسات حول انعكاسات جائحة كورونا على طلبة جنسيات أخرى ومقارنتها مع نتائج هذه الدراسة.

## المصادر والمراجع